\documentstyle[12pt]{article}
\begin{document}
\thispagestyle{empty}
\title{Exact Self-consistent Particle-like Solutions to the Equations 
of Nonlinear Scalar Electrodynamics in General Relativity}
\author{Yu. P. Rybakov and G. N. Shikin\\ 
Department of Theoretical Physics\\ 
Peoples' Friendship University of Russia\\ 
6, Miklukho-Maklay str., 117198 Moscow, Russia\\ 
e-mail: rybakov@udn.msk.su\\
B. Saha\\ 
Laboratory of Theoretical Physics\\ 
Joint Institute for Nuclear Research, Dubna\\ 
141980 Dubna, Moscow region, Russia\\ 
e-mail:  saha@thsun1.jinr.dubna.su} 
\date{}
\maketitle

\noindent
Exact self-consistent particle-like solutions with spherical and/or
cylindrical symmetry to the equations governing the interacting system 
of scalar, electromagnetic and gravitational fields have been obtained. 
As a particular case it is shown that the equations of motion admit a 
special kind of solutions with sharp boundary known as droplets. For 
these solutions, the physical fields vanish and the space-time is flat 
outside of the critical sphere or cylinder. Therefore, the mass and the 
electric charge of these configurations are zero. 


\newpage
\section{Introduction} 
\setcounter{equation}{0}    
Since the early history of elementary particle physics the attempts
to construct divergence-free theory had been being undertaken.
In 1912 G. Mie~\cite{mie} proposed the nonlinear modification of the
Maxwell equations, with the nonlinear electric current of the form
$j_{\mu} = (A_{\nu}A^{\nu})^2 A_{\mu}$. Within the scope of this
modification there do exist regular solutions approximizing the
electron structure.

In 1942 N. Rosen~\cite{ros} considered the system of interacting
electromagnetic and complex scalar fields  that also admitted the
existence of localized particle-like solutions. Nevertheless, these
two models suffered one and the same defect: the mass of the localized
object turned to be negative. Recently it was shown that this defect
of nonlinear electrodynamics can be corrected within the framework of
general relativity~\cite{chu}.

The aim of this paper is to consider self-consistent system of fields
to obtain particle-like configurations in the framework of general 
relativity. We show that in the case of electromagnetic scalar and 
gravitation fields system with specific type of interactions there 
exist droplet-like solutions having zero electric charge and mass. 
It is noteworthy to underline that the effective potentials, raised 
in this case, possess confining property i.e. create a strong repulsion 
on certain surfaces in configurational space.

\section{Fundamental Equations}  
\setcounter{equation}{0}

\noindent
As is known, 
there do not exist regular static spherically or cylindrically symmetric 
configurations within the framework of gauge invariant  nonlinear 
electrodynamics \cite{bron2}. One possible way to overcome this
difficulty is the nonlinear generalization of electrodynamics, 
with the use of the Lagrangian explicitly containing 4-potential $A_\mu$,  
\quad $\mu\,=\, 0,\ 1,\ 2,\ 3,$ thus breaking the gauge invariance inside 
the small critical sphere or cylinder. 
The introduction of the terms explicitly depending on potentials in 
electromagnetic equations presents the possibility to give an alternative
explanation of the processes like inelastic photon-photon interaction 
\cite{nov1}, galactic red-shift anomalies \cite{schiff}, \cite{peckev},
\cite{alfred}, electric screening at low temperature in the limit of
indirect interaction of photon with thermal neutrino background \cite{wolo}, 
the excess of high-energy photons in the isotropic flux \cite{lju}, 
avoiding the Big Bang singularity \cite{nov2}, origin of self-focused 
beam in the effective nonlinear vector field theory \cite{bissop}. 
The corresponding terms  appear in our scheme due  to  
the interaction between the electromagnetic and scalar  fields.  This 
interaction being negligible at large distances, the  Maxwellian 
structure of the electromagnetic equations (and  therefore  the gauge 
invariance) is reinstated far from the center of the system.

\noindent
We choose the Lagrangian in the form \cite{bron2}
\begin{equation}
{L}\,=\, \frac{R}{2\kappa} \,-\, \frac{1}{16\pi}\,
F_{\alpha \beta}\,{F}^{\alpha \beta}\,+\, \frac{1}{8\pi} \varphi_{,\alpha}\, 
{\varphi}^{,\alpha}\,\Psi(I),
\end{equation}
where $\kappa\,=\, 8\,\pi\,G$ is the Einstein's  
gravitational constant and the function $\Psi(I)$ of the invariant $I\,=\, 
{A}_\mu\,{A}^\mu$ characterizes the interaction between the 
scalar $\varphi$ and electromagnetic ${A}_\mu$ fields. In the sequel 
the function $\Psi(I)$ will be viewed as an arbitrary one, thus 
the Lagrangian (2.1) defines  the class  of models parameterized by 
$\Psi(I)$. In 1951 J. Schwinger \cite{Schw} used the special method 
to compute the effective coupling between a zero spin neutral meson and the 
electromagnetic field using some functions of electromagnetic 
field. Thus our approach to generate an effective Lagrangian generalizes 
the one proposed by Schwinnger. The particular choice of $\Psi (I)$  
will be made to obtain droplet-like configurations. 

\noindent
The field equations corresponding to the Lagrangian (2.1) read 
\begin{equation}
{\cal G}_{\mu}^{\nu}\,=\,-\,\kappa\,T_{\mu}^{\nu},
\end{equation}
\begin{equation}
\frac{1}{\sqrt{-g}} \frac{\partial}{\partial x^\alpha}
\bigl(\sqrt{-g}\,g^{\alpha \beta}\, \varphi_{,\beta} 
\, \Psi \bigr)\,=\,0, 
\end{equation}
\begin{equation}
\frac{1}{\sqrt{-g}} \frac{\partial}{\partial x^\beta}
\bigl(\sqrt{-g}\, {F}^{\alpha \beta} 
\bigr)\,-\, \bigl(\varphi_{,\beta}\,{\varphi}^{,\beta}\bigr)\, \Psi_I \, 
{A}^\alpha\,=\,0, 
\end{equation}  
where $\Psi_I\,=\, d\Psi/dI$ and ${\cal G}_{\mu}^{\nu} \,=\, { 
R}_{\mu}^{\nu}\, -\, \delta_{\mu}^{\nu}\, {R} /2 $ is the Einstein 
tensor. One can write the energy-momentum tensor of the interacting matter 
fields in the form:  
\begin{eqnarray}
T_{\mu}^{\nu}&=&(1/4 \pi)\,\bigl[\varphi_{,\mu}\,\varphi^{,\nu}\, 
\Psi(I)\,-\,{F}_{\mu \alpha}\,{F}^{\nu \alpha}\,+\,
\varphi_{,\alpha}\, \varphi^{,\alpha}\,\Psi_I\,{A}_\mu\,{
A}^\nu \bigr] \nonumber \\
&-& \delta_{\mu}^{\nu}
\bigl[\frac{1}{8\pi} \varphi_{,\beta}\,{\varphi}^{,\beta}\, \Psi(I)
\,-\,\frac{1}{16\pi}\,{F}_{\alpha \beta}\,{F}^{\alpha \beta}\bigr]. 
\end{eqnarray}

\section{Configurations with spherical symmetry}
\setcounter{equation}{0}
Searching for the static  spherically-symmetric  solutions  to 
the system of equations (2.2)\,-\,(2.4) , we consider the metric  in  the 
form \cite{bron3}:
\begin{equation}
ds^2\,=\,e^{2\gamma}\,dt^2\,-\,e^{2\alpha}\,d\xi^2\,-\, 
e^{2\beta}\bigl[d\theta^2\,+\,\mbox{sin}^2\theta\ d\phi^2 \bigr], 
\end{equation}
with $\xi$ being the radial variable.
Let us now formulate the requirements to be fulfilled by particle-like 
solutions (PLS). These are \cite{bron4}

\noindent
(a) Stationarity [applied to the metric (3.1)] i.e.
$$ \alpha\,=\,\alpha(\xi), \quad \beta\,=\,\beta(\xi), \quad \gamma\,=\,
\gamma(\xi);$$

\noindent
(b) regularity of the metric and the matter fields in the whole 
space-time;

\noindent
(c) asymptotically Schwarzschild metric and corresponding behavior of the 
field functions.

\noindent
In view of requirement (a) it is convenient to choose the harmonic $\xi$
coordinate ($\Box \xi\,=\,0$) in (3.1) to satisfy the subsidiary condition 
\cite{bron5}:  
\begin{equation} \alpha\,=\,2\,\beta\,+\,\gamma.  \end{equation} 
The corresponding coordinate in flat space-time is just $\xi\,=\,1/r.$
With the constraint (3.2) the system of Einstein equations (2.2) reads:  
\begin{equation}
e^{-2 \alpha}\,(2 \beta^{\prime \prime}\,-\, U)\,-\, e^{-2 \beta}\,=\, 
-\kappa\,T_{0}^{0},
\end{equation}
\begin{equation}
e^{-2 \alpha}\, U \,-\, e^{-2 \beta}\,=\, 
-\kappa\,T_{1}^{1},
\end{equation}
\begin{equation}
e^{-2 \alpha}\,(\beta^{\prime \prime}\,+\,\gamma^{\prime \prime}\,-\, 
U)\,=\, -\kappa\,T_{2}^{2}\,=\,-\kappa\,T_{3}^{3},
\end{equation}
where $U\,=\, \beta^{\prime 2}\,+\, 2\,\beta^{\prime}\,\gamma^{\prime}$,
and prime $(^\prime)$ denotes differentiation with respect to $x$.
Note that the field functions, as well as 
the components of  the metric tensor depend  on the single 
spatial variable $\xi$.  Assuming the electromagnetic field to be 
determined by the time  component ${A}_0 \,=\,{A}(\xi)$ of the  
4-potential one finds the  unique non-trivial component of the field  
tensor  ${F}_{10} \,=\, {A}^{\prime},$ and  the invariant $I$ 
reduces to $I\,=\, e^{-2\gamma}\,{A}^2(\xi).$  

One can write the  non-zero  components  of  the  energy-momentum 
tensor (2.5) as follows:
\begin{equation}
T_{0}^{0}\,=\,(1/8\pi)\,e^{-2\alpha}\,\bigl[{A}^{\prime 
2}\,e^{-2\gamma} \,+\, \varphi^{\prime 2}\bigl(\Psi\,-\,2\,{A}^2\, 
e^{-2\gamma}\,\Psi_I\,\bigr)\bigr],
\end{equation} 
\begin{equation}
T_{1}^{1}\,=\,-\,T_{2}^{2}\,=\,-\,T_{3}^{3}\,=
\,(1/8\pi)\,e^{-2\alpha}\,\bigl[{A}^{\prime 
2}\,e^{-2\gamma} \,-\, \varphi^{\prime 2}\,\Psi\bigr].
\end{equation} 
Adding together the equations (3.4) and (3.5) and using the property
$T_{1}^{1}\,+\,T_{2}^{2}\,=\, 0$, one obtains the differential equation
$$\beta^{\prime \prime}\,+\,\gamma^{\prime 
\prime}\,-\,e^{2(\beta\,+\,\gamma)}\,=\,0,$$                       
with the solution \cite{bron6}
\begin{equation}
e^{-(\beta\,+\,\gamma)}\,=\,{\cal 
S}(k,\,\xi)\,=\,\left\{\begin{array}{ccc} k^{-1}\,\mbox{sh}\,k\xi,& 
k\,>\,0, \\ \xi,& k\,=\,0, \\ k^{-1}\, \mbox{sin}\,k\xi,& k\,<\,0, 
\end{array}\right .
\end{equation}
depending on the 
constant $k$. Notice  that  another  constant  of integration is trivial, 
so that $\xi \,=\, 0$ corresponds to the  spatial infinity, where 
$e^\gamma \,=\,1$ and $e^\beta \,=\,\infty$.  Without loss of generality 
one can choose $\xi \,>\, 0.$  

The scalar field equation (2.3) has the evident solution 
\begin{equation}
\varphi^{\prime}\,=\,C\,P(I),
\end{equation} 
where $P(I)\,=\, 1/\Psi(I)$ and $C$ is  the  
integration constant.  Putting (3.9) into (2.4) one gets the equation for 
the electromagnetic field 
\begin{equation}
\bigl(e^{-2\gamma}\,{A}^{\prime}\bigr)^{\prime}\,-\, C^2\, P_I\, 
e^{-2\gamma}\,{A}\,=\,0,
\end{equation} 
where the second term could  be naturally interpreted  as  the 
induced nonlinearity.  
In view of (3.9) one rewrites the Einstein equation 
(3.4) and the result of adding the equations (3.3) and (3.4) as follows :  
\begin{equation}
\gamma^{\prime 2}\,=\,-G\,\bigl(C^2\,P\,-\,{A}^{\prime 
2}\,e^{-2\gamma} \bigr)\,+\,K, \qquad  K\,=\, k^2\,\mbox{sign}k, 
\end{equation}
\begin{equation}
\gamma^{\prime \prime}\,=\,G\,e^{-2\gamma}\,\bigl({A}^{\prime 2}\, 
+\,C^2\,{A}^2\,P_I \bigr).
\end{equation} 
One can easily check that the 
equation (3.11) is the first integral of the equations (3.10) and (3.12).  
Eliminating the term $(P_I\,{A})$ between (3.10) and (3.12) one gets 
the equation:  
\begin{equation}
\gamma^{\prime \prime}\,=\,G\,\bigl({A}\,{
A}^{\prime}\,e^{-2\gamma}\bigr)^{\prime}, \end{equation} 
with the evident first integral: 
\begin{equation}
\gamma^{\prime}\,=\,G\,{A}\,{
A}^{\prime}\,e^{-2\gamma}\,+\,C_1, \quad C_1\,=\,\mbox{const.} 
\end{equation} 
Let us consider the simple case $C_1 \,=\,0.$ Then from 
(3.14) we get 
\begin{equation} 
e^{2\gamma}\,=\,G\,{A}^2\,+\,H,\qquad H\,=\,\mbox{const.}  
\end{equation} 
Substituting $\gamma^{\prime}$ and 
$e^{2\gamma}$ from (3.14) and (3.15) into (3.10), we find for ${
A}(\xi)$ the differential equation:  
\begin{equation} {A}^{\prime 
2}\,(G\,{A}^2\,+\,H)^{-2}\,=\, (G\,C^2\,P\,-\,K)/\,G\,H, 
\end{equation}  
which can be solved by quadrature:  
\begin{equation}
\int\limits_{}^{}\frac{d {A}}{(G\,{A}^2\,+\,H)\, 
\sqrt{G\,C^2\,P\,-\,K}}\,=\,\pm\,(1/\sqrt{G\,H})\, (\xi\,-\,\xi_0), \quad
\xi_0\,=\,\mbox{const.}
\end{equation} 
It is clear that the 
configuration obtained has a center if and only if $e^{\beta} \,=\, 0$ at 
some $\xi\,=\, \xi_c.$  One  can  show \cite{bron5} that the conditions for 
the center $\xi_c \,=\, \infty$ to be regular imply $K\,=\,0$ and  the 
following behavior of the field quantities in the vicinity of the point 
$\xi_c\,=\,\infty$:
\begin{eqnarray}
\gamma^{\prime}\,=\,O\,(\xi^{-2}), \quad {A}^{\prime} \to 
{A}_c\,\ne\,\infty, \quad {A}^{\prime}\to 0, \nonumber\\
\xi^4\,P(I) \to 0, \quad \mid \xi^4\,I\,P_I \mid\,<\,\infty.
\end{eqnarray}
In view of (3.18) we deduce from (3.14) that $C_1 \,=\,0$ in accordance 
with the earlier supposition.   

\noindent
Now we can write the boundary conditions on 
the surface  of the critical sphere $\xi\,=\,\xi_0$:
\begin{equation}
T_{\mu}^{\nu}\,=\,{A}\,=\,{A}^{\prime}\,=\,0, \quad e^{\gamma}\, 
=\,1,\quad e^{\beta}\,=\, 1/\xi_0\,>\,0.
\end{equation}
Due to (3.19) and (3.15) we infer that $H=1.$  The condition $K=0$ leads
to $k=0$ in (3.8) and the space-time (3.1) that fulfills the regularity
conditions (3.18) takes the form
\begin{eqnarray}
ds^2\,=\,(G{A}^2 + 1) dt^2 - \frac{1}{\xi^2\,(G{A}^2 + 1)}
\Bigl(\frac{d\xi^2}{\xi^2} + [d\theta^2 + \mbox{sin}^2 \theta d\phi^2]
\Bigr). 
\end{eqnarray}
We can finally write ${A}$ and $\varphi$ as follows:
\begin{equation}
\int\limits_{}^{}\frac{d {A}}{(G\,{A}^2\,+\,1)\, 
\sqrt{P}}\,=\,\pm\, C\,(\xi\,-\,\xi_0), 
\end{equation} 
\begin{equation}
\varphi = C \int P d\xi = \int\, \sqrt{P} e^{-2\gamma}d{A} 
= \int\,\frac{\sqrt{P}d{A}}{G{A}^2 + 1}. 
\end{equation}
Let us now calculate the matter field energy density: 
\begin{equation}
T_{0}^{0}\,=\,(C^2/\,8\,\pi)\,e^{-2\alpha}\,\bigl[\,P\,(1\,+\,e^{2\gamma} 
)\,+\,2\,I\,P_I(I)\,\bigr].
\end{equation} 
One can readily derive from (3.23) the energy $E_f$ of the matter fields:  
\begin{equation}
E_f\,=\,\int\limits_{}^{}\,d^3 x\,\sqrt{-^3\,g}\,T_{0}^{0}\,=\,
(C/2)\,\int\limits_{A(\xi=0)}^{A(\xi \to \infty)}\,d{A}\,e^{-3\gamma}\, 
\bigl[\sqrt{P}(1\,+\,e^{2\gamma})\,+\,4\,I\,(\sqrt{P})_I\,\bigr]. 
\end{equation} 
Thus the equations to the scalar and electromagnetic fields are
completely integrated. As one sees, to write the scalar $(\varphi)$ 
and vector $(A)$ functions, as well as the energy density $(T_{0}^{0})$
and energy of the material fields $(E_f)$, explicitly, one has to give  
$P(I)$ in explicit form. Here we will give the detail analysis for
some concrete forms of $P(I)$.

\noindent
{\bf I.} Let us consider $P(I)$ in the form
\begin{equation}
P(I) = P_0 (\lambda I - N)^s\,R(\lambda I), \qquad 2\le s \le 3,
\end{equation}
where  $R(\lambda I)$ is some arbitrary, continuous, positive defined
function, having non-trivial value at the center; $\lambda$ is the 
coupling parameter; $N > 0$ is some dimensionless constant that is
equal to the value of $\lambda I$ at the center. The other constant
$P_0$ is defined from the condition $P = 1$ at spatial infinity
$\xi = 0$. For $R =$ const. one gets the most simple form of $P(I)$
that leads to regular solutions. In this case the energy density is positive
if $\lambda I \ge N.$ 

{\bf a)} Choosing  $P(I)$ in the form
\begin{equation}
P(I) = P_0 (\lambda I - N)^2,
\end{equation}
we get
\begin{equation}
{A}(\xi) = \sqrt{\frac{N}{\lambda - GN}} \mbox{cth}\Lambda (\xi + \xi_1),
\end{equation}
where $\Lambda = \sqrt{C^2 N P_0 (\lambda - GN)}$, the integration
constant $\xi_1$ is defined from ${A}(0) = m/q$ with $m$ and $q$ 
being the mass and the charge of the system. In this case we get
$$ P_0 = \bigl(\lambda m^2 /q^2 - N\bigr)^{-2}, \qquad 
\lambda m^2 /q^2 > N.$$
Inasmuch $\sqrt{\lambda} m/ |q| > \sqrt{N}$, then 
$\delta = \sqrt{G} m/|q| > \sqrt{GN/\lambda} =\sigma$. Taking $\delta < 1$
and $\sigma < 1$ we get the inequality:
$$ 0 < \sigma < \delta < 1.$$
Now we can rewrite $P_0$ in the form 
$$ P_0 = \frac{G^2}{\lambda^2}\,(\delta^2 - \sigma^2)^{-2}.$$
The metric function $e^{2\gamma}$, electric field and the total energy
of the material field system can be written as 
\begin{equation}
e^{2\gamma} = G{A}^2 + 1 = \frac{C^2}{q^2}\Bigl[\frac{\sigma^2}
{1 - \sigma^2}\mbox{cth}^2 \Lambda (\xi + \xi_1) + 1\Bigr],
\end{equation}
\begin{equation}
|{\bf E}| = (-{F}_{10}{F}^{10})^{1/2} =
\frac{\Lambda \sqrt{N}}{\sqrt{\lambda (1 - \sigma^2)}} \frac{\xi^2}
{\mbox{sh}^2 \Lambda (\xi + \xi_1)},
\end{equation}
\begin{eqnarray}
E_f &=& \frac{q}{2\sqrt{G}}\Bigl[\frac{\delta - \sigma}{\delta + \sigma}\,
\frac{\delta + 2\sigma}{3} + \frac{4(\delta^2 + \delta\sigma +
\sigma^2) - 3}{3(\delta  + \sigma)} \nonumber\\
&+& \frac{1-\sigma^2}{2(\delta^2 - \sigma^2)}\mbox{ln}\,
\frac{(1 + \delta)(1 -\sigma)}{(1 - \delta)(1 + \sigma)} \Bigr].
\end{eqnarray}
As one sees
$$ E_f|_{\delta \to \sigma} \to \frac{q\delta}{\sqrt{G}} = m, \qquad
E_f|_{\delta \to 1} \to \infty.$$
The infinite vale of $E_f$ can be interpreted as the physical reason of
existence of limitation $\delta < 1.$

\noindent
{\bf b)} Let us consider the case with $I_c = 0$, choosing
\begin{equation}
P(I) = \lambda I.
\end{equation}
On the spatial infinity, where $ I = I_0 = m^2 /q^2$, $P = 1$, that leads to
$\lambda = q^2/m^2$, i.e. the coupling constant is connected with mass
and charge. In this case we get
\begin{equation}
{A}(\xi) = \frac{1}{\sqrt{G}\mbox{sh}\, m(\xi + \xi_1)},
\end{equation}
where as in previous case $\xi_1$ is defined from ${A}(0) = m/q$.
The metric function $e^{2\gamma}$, electric field and the total energy
of the material field system can be written as 
\begin{equation}
e^{2\gamma} =  \frac{C^2}{q^2}
\mbox{cth}^2 \,mC(\xi + \xi_1)/q,
\end{equation}
\begin{equation}
|{\bf E}| = \frac{mC^2}{q^2 \sqrt{G}}\frac{\xi^2 \mbox{ch}\,mC(\xi +\xi_1)/q}
{\mbox{sh}^2\,mC(\xi +\xi_1)/q},
\end{equation}
\begin{eqnarray}
E_f = \frac{q}{4\sqrt{G}}\Bigl[3\delta \frac{1}{\delta}\mbox{ln}\,
(1 - \delta^2)\Bigr].
\end{eqnarray}
As one sees
$$ E_f|_{\delta \ll 1} \approx m, \qquad
E_f|_{\delta \to 1} \to \infty.$$

\noindent
{\bf II.} 
A specific type of solutions to the nonlinear field equations in flat 
space-time were obtained in a series of interesting articles 
\cite{wer1}, \cite{wer2}, \cite{wer3}, \cite{wer4}. These 
solutions are known as droplet-like solutions or simply droplets. 
Distinguishable  property of these solutions is the availability of some 
sharp boundary, defining the space domain, in which the material 
field happens to be located i.e. the field is zero beyond this area. As 
was found the solutions mentioned exist in field theory with specific 
interactions that can be considered as effective, generated by initial 
interactions of the unknown origin. Contrary to  the  widely 
known soliton-like solutions, with field  functions  and  energy 
density asymptotically tending to zero at spatial infinity,  the 
solutions in question vanish at a finite distance from the center 
of the system (in the case of spherical  symmetry)  or  from  the 
axis (in the case of cylindrical symmetry).  Thus, there  exists 
the sphere or cylinder with critical radius $r_0$, outside of which 
the fields disappear. Therefore the field configurations have the 
droplet-like structure \cite{wer1}, \cite{bron1}, \cite{ryb1}. 

\noindent
Let us now choose the function $P(I)$ as follows \cite{ryb2}
[see Figure 2]:  
\begin{equation}
P(J)\,=\,J^{(1\,-\,2/\sigma)}\,\bigl[(1\,-\,J)^{1/\sigma}\,-\, 
J^{1/\sigma} \bigr]^2\,(1\,-\,J),
\end{equation}  
where $J\,=\, G\,I;\quad \sigma\,=\, 2n\,+\,1;\quad  n=1,2,3\cdots$  
Then on account of $K\,=\, 0$ and $H\,=\,1$ we get from (3.17)  the 
following expression for ${A}(\xi)$ [see Figure 1]:  
\begin{equation}
A (\xi \le \xi_0) = 0, \quad
{A}(\xi \ge \xi_0)\,=\,(1/\sqrt{G})\,\bigl[1\,-\, 
\mbox{exp}\,\bigl(-\frac{2\,C\,\sqrt{G}}{\sigma}\,(\xi\,-\,\xi_0)\bigr) 
\bigr]^{\sigma/2}.
\end{equation}  
As one can see from (3.37), the 
conditions (3.18) for the center to be regular and the 
matching conditions (3.19) on  the surface of the critical sphere are 
fulfilled if $\sigma\,>\,2.$ It is also obvious from (3.37) that for $\xi 
< \xi_0$ the value of square bracket turns to be negative one and ${
A}(\xi)$ becomes imaginary since $\sigma$ is an odd number. Since we are 
interested in the real ${A}(\xi)$ only, without loss of generality 
we may assume the value of ${A}(\xi)$ be zero for $\xi \le \xi_0$, the
matching at $\xi = \xi_0$ being smooth. 

\noindent
Recalling that $J\,=\,G\,{A}^2 /\,(G\,{A}^2 \,+\,1)$, we get 
from (3.37) that $J(\infty)\,=\, 1/2$ and $J(\xi_0)\,=\,0$, thus implying: 
\begin{equation}
P(I)\mid_{\xi\,=\,\infty} \,=\,P(I)\mid_{\xi\,=\,\xi_0}\,=\,0.
\end{equation}
It means that at $\xi\,=\,\xi_c\,=\,\infty$ and $\xi\,=\,\xi_0$, the 
interaction function $\Psi(I)\,=\,1/P(I)$ is singular.  It  turns out 
nevertheless that  the energy density $T_{0}^{0}$  is regular at these 
points due to the fact that it contains $\Psi(I)$ as a multiplier in the 
form:  
\begin{equation}
e^{-2\alpha}\,\varphi^{\prime 2}\,\Psi\,=\,C^2\,e^{-2 \alpha}\,P(I),
\end{equation}
which tends to zero as $\xi\,\to\,\xi_c$ or $\xi\,\to\,\xi_0.$  
As follows from (3.37), for the limiting case $\xi_0\,=\,0$, when the 
critical sphere goes to the spatial infinity and the solution in question 
is defined at $0\,\le\,\xi\,\le\,\infty$, it appears that  at spatial 
infinity $(\xi\,=\,0)\quad {A}\,=\,0$ and $P(I)\,=\,0$. In this case 
we obtain the usual soliton-like configuration not possessing any sharp 
boundary.  

\noindent
It should be emphasized that at spatial infinity $(\xi\,=\,0)$ one can 
compare the metric 
found with the Schwarzschild one and the electrical field with the Coulomb 
one, thus determining the total mass $m$ and the charge $q$ of the system:  
$$G\,m\,=\, -\,\gamma^{\prime}(0),\quad q\,=\,-\,{A}^{\prime}(0).$$ 
Taking into account that $e^{2\gamma}\,=\,G\,{A}^2 \,+\,1$, one can 
find through the use of (3.37) that for $\xi_0\,=\,0,\quad {
A}^{\prime}(0)\,=\,-q\,=\,0$ and $\gamma^{\prime}(0)\,=\,-G\,m\,=\,0.$  
Therefore, the total energy of the soliton-like system,defined as the sum 
of  the material fields energy and that of the gravitational field, 
vanishes.  If now one chooses the integration constant $\xi_0\,>\,0$,  
then the field configuration with the sharp boundary (droplet) appears. In 
this case for $\xi\,\le\,\xi_0$  one obtains ${A}(\xi)\,=\,0$ and  
$e^{2\gamma}\,=\,1$, i.e.  outside of the droplet gravitational and 
electromagnetic fields disappear, that implies the vanishing  of the total 
mass  and  the charge of  the system.  This unusual property makes the 
droplet-like object poorly visible  for the outer observer.  
   
\noindent
It should be emphasized that the field 
energy is localized in the region $(\xi_0 \,\le\,\xi\,<\,\infty):$
\begin{equation}
T_{0}^{0}(\xi)\mid_{\xi\to \infty}\,\to \,0, \qquad
T_{0}^{0}(\xi)\mid_{\xi\to \xi_0}\,\to \,0, 
\end{equation} 
namely, inside the critical sphere with the 
radius 
$$ R \,=\, \int\limits_{0}^{\infty}\,d\xi\,e^{\alpha(\xi)}
\,=\, \int\limits_{0}^{\infty}\,d\xi\,/\xi^2\Bigl\{\bigl[1 - e^
{-2C\sqrt{G}(\xi-\xi_0)/\sigma}\bigr]^\sigma +1\Bigr\}^{(1/2)}
\,< \infty .$$  
Taking into account that $ e^{2\gamma}\,=\,1/(1\,-\,J)$ and  
$e^{-3\gamma}\,d{A}\,=\,dJ/2\sqrt{G\,J},$ we rewrite total energy
of the material fields in terms of $J$:  
\begin{eqnarray} 
E_f\,=\,(C/4\sqrt{G})\,\int\limits_{0}^{1/2} \Bigl\{4 \frac{d\sqrt{J\,P}}
{dJ}\,+\,\frac{\sqrt{P\,J}}{1-J}\Bigr\}\,dJ. \nonumber
\end{eqnarray}
Contribution of the first term of the foregoing equality is trivial for the 
choice of $P(I)$ in the form (3.36) as in this case 
$P(I)|_{0}=P(I)|_{\lambda/2}=0$. As $P(I)$ is positive and $J$ lies in the
interval $(0,\,1/2)$, one estimates 
\begin{eqnarray}
E_f = \frac{C}{4\sqrt{G}}\,\int\limits_{0}^{1/2}\,
\frac{\sqrt{P\,J}}{1-J}\,dJ> 0. \nonumber
\end{eqnarray}
Let us note that we consider the constant $C$ to be a positive one.
Knowing that the total energy of the droplet-like  
object is zero this inequality implies the negativity of its gravitational 
energy. Thus the droplet-like configuration of the fields obtained is 
totally regular with zero total energy (including the energy of 
proper gravitational field) and null electric charge and remains 
unobservable to one located outside the sphere with radius $R$ 
\cite{ryb2}, \cite{ryb3}. 
In order to clarify the fact that the role of the gravitational 
field in forming the droplet-like configuration is not decisive it is 
worthwhile to compare the solution obtained with that in the flat 
space-time, described by the interval 
$$ ds^2\,=\,dt^2\,-\, dr^2\,-\,r^2\,[d\theta^2\,+\, \mbox{sin}^2\theta\, 
d\phi^2].$$  
In the latter case the equation (2.3) 
admits the solution 
\begin{equation}
\varphi^{\prime}(r)\,=\,- C\,P(I)/r^2.
\end{equation}
Substituting (3.41) into (2.4), 
one finds that the equation for the electromagnetic field can be solved by 
quadrature:  
\begin{equation}
\int\limits_{}^{} d{A}/\sqrt{P}\,=\,\pm\,C\,\bigl(\frac{1}{r}\,-\, 
\frac{1}{r_0}\bigr), \quad r_0\,=\,\mbox{const.}
\end{equation}   
Note that the droplet-like configuration ${A}(r)$ will be similar to 
(3.37) if one chooses the function $P(I)$ more simple than (3.36):  
\begin{equation}
P(I)\,=\,J^{1\,-\,2/\sigma}\,(1\,-\,J^{1/\sigma})^2, \qquad 
J\,=\,\lambda\,I,\end{equation}  
where $\lambda\,=\,$ const;\quad $\sigma\,=\,2\,n\,+\,1;\quad 
n\,=\,1,2,3,\cdots$. Then substituting (3.43) into (3.42) one gets the 
solution 
\begin{equation}
{
A}(r)\,=\,(1/\sqrt{\lambda})\,\bigl[1\,-\,\mbox{exp}\bigl(-\frac{2\,C\,
\lambda}{\sigma}\,(\frac{1}{r}\,-\,\frac{1}{r_0})\bigr)\bigr]^{\sigma/2}. 
\end{equation} 
One can see from (3.44) that ${A}(r)\,=\,0$ as $r\,\ge\,r_0$, i.e. 
the charge of the flat space-time droplet configuration also vanishes.  
For this solution the regularity conditions at the center $r\,=\,0$ and  
on the surface of the critical sphere $r\,=\,r_0$  are evidently 
fulfilled.  It similarly appears that for $r\,=\,\infty$  one finds the 
usual soliton-like structure with field vanishing as $r\,\to \infty$.  
The field energy $E_f$ is defined as follows:  
\begin{equation}
E_f\,=\,C\,\int\limits_{{A}(r_0)}^{{A}(0)}\,d{A}\, 
(\sqrt{P} \,+\,I\,P_I/\sqrt{P})\,=\,C\sqrt{P\,I}\mid_{{
A}(r_0)}^{{A}(0)}.  
\end{equation}  
Inspecting that $P\,I\,=\,0$ both at $r\,=\,0$ and $r\,=\,r_0$, we arrive 
through (3.45) at $E_f \,=\,0.$  

\noindent
Thus in the flat space-time as well as 
for the self-gravitating system, the total energy and charge of the 
droplet-like configuration vanish.  

\section{Configurations with cylindrical symmetry}
\setcounter{equation}{0}                                  

\noindent
Obviously, in view of physics, the most interesting case is the spherically
symmetric one, nevertheless in some cases it is necessary to study the 
two-dimensional cylindrically symmetric regular solutions in the vicinity
of symmetry axis (vortex~\cite{niel}, string-like solutions~\cite{terlet}).
These solutions can describe realistic objects like fluxion~\cite{abr},
light-beam~\cite{zakha} and can serve as the logical approximation to
the objects with toroidal structure~\cite{vega}. 
Let us now search for static cylindrically-symmetric solutions to 
the equations (2.2)-(2.4). In this case the metric can  be chosen as 
follows \cite{bron7}, \cite{shik1}:  
\begin{equation}
ds^2\,=\,e^{2\gamma}\,dt^2\,-\,e^{2\alpha}\,dx^2\,-\, 
e^{2\beta}\,d\phi^2\,-\,e^{2\mu}\,dz^2. 
\end{equation}
The requirements to be fulfilled by soliton-like 
solutions in this case are \cite{shik2}

\noindent
(a) Stationarity [applied to the metric (4.1)] i.e.
$$ \alpha\,=\,\alpha(x), \quad \beta\,=\,\beta(x), \quad \gamma\,=\,
\gamma(x), \quad \mu\,=\,\mu(x).$$
It means for (4.1) that all the components 
of the metrical tensor depend on the single spatial coordinate $x\, \in\, 
[x_0, \,x_a],$ where $x_a$ is the value of $x$ on the axis of symmetry, 
defined by the condition $\mbox{exp}[\beta(x_a)]\,=\,0$, and $x_0$ is the 
value of $x$ on the surface of the critical cylinder.  The coordinates $z$ 
and $\phi$ take their standard values: $z\,\in\,[-\infty,\,\infty], \quad 
\phi\,\in\, [0,2\pi].$  

\noindent
(b) regularity of the metric and the matter fields in the whole 
space-time;

\noindent
(c) localized in space-time (with finite field energy):
$$E_f\,=\,\int\,T_{0}^{0}\,\sqrt{-^3 g}\ dV \,<\, \infty.$$ 

\noindent
Requirement (c) assumes the rapid decreasing of energy density of material 
field at spatial infinity, which together with (b) guaranties the 
finiteness of $E_f$. Let us note that $E_f$ may be finite even for
singular solutions on the axis. Requirement (b) means the regularity of 
material fields as well as the regularity of metric functions that entails 
the demand of finiteness of energy-momentum tensor of material fields all 
over the space. 
If the system considered contains scalar $\varphi$ and electric ${\bf E}$
(or magnetic ${\bf H}$) fields, the regularity conditions on $x=x_a$
take the form~\cite{bron7}:
\begin{eqnarray}
e^{\beta} = 0; \quad |\gamma| < \infty; \quad |\mu| < \infty; \quad
e^{2(\beta - \alpha)}(\beta^\prime)^2 = 1; \quad e^{-2\alpha} 
(\gamma^\prime)^2 = 0; \nonumber \\
\{|{\bf E}| = 0; \quad |{\bf H_{\parallel}}| < \infty; \quad
|{\bf H_{\bot}}| = 0 \}; \quad |T_{\mu}^{\nu}| < \infty,
\end{eqnarray}
where ${\bf H_{\parallel}}$ and  ${\bf H_{\bot}}$ are the longitudinal
and transverse magnetic fields defined as chronometric invariants
~\cite{Mits}. In view of requirement (a) it is convenient to choose 
the coordinate $x$ in (4.1) to satisfy the subsidiary condition 
\cite{shik1}:  
$$ \alpha\,=\,\beta\,+\,\gamma\,+\,\mu, $$  
that permits to present the system of the Einstein equations  in the form: 
\begin{equation}
\mu^{\prime \prime}\,+\,\beta^{\prime \prime}\,-\, V\,=\,
-\,\kappa\,T_{0}^{0}\,e^{2 \alpha}, \end{equation} 
\begin{equation}
\mu^{\prime}\,\beta^{\prime}\,+\,\beta^{\prime}\,\gamma^{\prime}\,+
\,\gamma^{\prime}\,\mu^{\prime}\,=\,V\,=\,-\kappa\,T_{1}^{1}\,e^{2\alpha}, 
\end{equation} 
\begin{equation}
\gamma^{\prime \prime}\,+\,\beta^{\prime \prime}\,-\, V\,=\,
-\,\kappa\,T_{2}^{2}\,e^{2 \alpha}, \end{equation} 
\begin{equation}
\mu^{\prime \prime}\,+\,\gamma^{\prime \prime}\,-\, V\,=\,
-\,\kappa\,T_{3}^{3}\,e^{2 \alpha}. \end{equation} 
As in the preceding section, the electromagnetic 
field is described by the time component of the 4-potential ${A}_0(x) 
\,=\,{A}(x)$ and by the component ${F}_{1\,0}\,=\,d{
A}/dx\,=\,{A}^{\prime}$ of the field strength tensor and the 
energy-momentum tensor of interacting fields is defined by the 
equations (3.6), (3.7). 

\noindent
Adding together the equations (4.4) and (4.5) and 
using  (3.7), one obtains the simple equation: 
\begin{equation}
\gamma^{\prime \prime}\,+\,\beta^{\prime \prime}\,=\,0,
\end{equation} 
with the solution 
\begin{equation}
\beta(x)\,+\,\gamma(x)\,=\,C_2\,x,\qquad C_2\,=\,\mbox{const.}
\end{equation}
Notice that the second 
integration constant in (4.8) can be taken trivial, as it determines only 
the choice of scale.  

\noindent
In a similar way the addition of equations (4.4) and 
(4.6) leads to the equation:
\begin{equation}
\gamma^{\prime \prime}\,+\,\mu^{\prime \prime}\,=\,0,
\end{equation} 
with the solution 
\begin{equation}
\mu(x)\,+\,\gamma(x)\,=\,C_3\,x,\qquad C_3\,=\,\mbox{const.}
\end{equation}
whereas the substraction of (4.5) and (4.6) gives 
\begin{equation}
\beta^{\prime \prime}\,-\,\mu^{\prime \prime}\,=\,0,
\end{equation} 
with the solution 
\begin{equation}
\beta(x)\, - \,\mu(x)\,=\,C_4\,x,\qquad C_4\,=\,\mbox{const.}
\end{equation}
Solving the equation (2.2) in the metric (4.1), 
one gets  the same result as in (3.9), i.e. 
\begin{equation}
\varphi^{\prime}(x)\,=\,C\,P(I).
\end{equation} 
Substituting (4.13) into (2.4),  one  finds the equation for the 
electromagnetic field, coincident with (3.10) i.e.
\begin{equation}
\bigl(e^{-2\gamma}\,{A}^{\prime}\bigr)^{\prime}\,-\, C^2\, P_I\, 
e^{-2\gamma}\,{A}\,=\,0,
\end{equation} 
where the second term could  be naturally interpreted  as  the 
induced nonlinearity.  
Now as in the previous 
case, we use the equation (4.4) and  sum of equations (4.3) and (4.4) 
which in view of (4.8) and (4.10), take the form: 
\begin{equation}
\gamma^{\prime 2}\,-\,C_2\,C_3\,=\,-G\,\bigl(C^2\,P\,-\,{A}^{\prime 
2}\,e^{-2\gamma} \bigr), 
\end{equation}
\begin{equation}
\gamma^{\prime \prime}\,=\,G\,e^{-2\gamma}\,\bigl({A}^{\prime 2}\, 
+\,C^2\,{A}^2\,P_I \bigr).
\end{equation} 
Elimination of $P_I\,{A}$ between the 
equations (4.14) and (4.16) gives  the equation 
\begin{equation}
\gamma^{\prime \prime}\,=\,G\,\bigl({A}\,{
A}^{\prime}\,e^{-2\gamma}\bigr)^{\prime}, \end{equation} 
with the evident first integral: 
\begin{equation}
\gamma^{\prime}\,=\,G\,{A}\,{
A}^{\prime}\,e^{-2\gamma}\,+\,C_1, \quad C_1\,=\,\mbox{const.} 
\end{equation} 
Integrating (4.18) under the choice $C_1\,=\,0,$ one again obtains  
\begin{equation} 
e^{2\gamma}\,=\,G\,{A}^2\,+\,H,\qquad H\,=\,\mbox{const.}  
\end{equation} 
Finally, substituting $\gamma^{\prime}$ from (4.18) and 
$e^{2\gamma}$ from (4.19) into (4.15), one gets the equation for 
${A}(x):$ 
\begin{equation} {A}^{\prime 
2}\,(G\,{A}^2\,+\,H)^{-2}\,=\, (G\,C^2\,P\,-\,C_2\,C_3)/\,G\,H. 
\end{equation}  
The equation (4.20) can be solved by quadrature:
\begin{equation}
\int\limits_{}^{}\frac{d {A}}{(G\,{A}^2\,+\,H)\, 
\sqrt{G\,C^2\,P\,-\,C_2\,C_3}}\,=\,\pm\,(1/\sqrt{G\,H})\, 
(x\,-\,x_0). 
\end{equation} 

\noindent
Let us formulate regularity conditions to be satisfied by the 
solutions to the equations (2.2)-(2.4) on the  axis  of symmetry defined 
by the value $x\,=\,x_a,$ where $\mbox{exp}[\beta(x_a)]\,=\,0.$  
As according to the regularity conditions formulated earlier
$|\gamma (x_a)| < \infty$ and $|\beta (x_a)| < \infty$ from (4.8) and (4.12)
one gets $\beta (x) \approx C_2 x \to -\infty$ (whereas $x_a = -\infty$ if
$C_2 > 0$  and $x_a = +\infty$ if $C_2 < 0$); 
$\beta (x) \approx C_4 x \to -\infty$ (whereas $x_a = -\infty$ if
$C_4 > 0$  and $x_a = +\infty$ if $C_4 < 0$). It leads to $C_2 = C_4$,\,\,
$\gamma(x) \equiv -\mu(x)$ and $\alpha(x) \equiv \beta(x)$. As one sees,
from $\gamma(x) \equiv -\mu(x)$ follows $C_3 = 0.$ The regularity 
conditions are similar to (3.18) for the case of spherical 
symmetry, implying that the following relations hold as $x \to x_a \,=\, 
\infty:$
\begin{eqnarray}
\gamma^{\prime}\,\to\,0, \quad {A}^{\prime} \to 
{A}_c\,\ne\,\infty, \quad {A}^{\prime}\to 0, \nonumber\\
e^{2\mid C_2\mid x}\,P(I) \to 0, \quad 
e^{2\mid C_2\mid x}\,\mid I\,P_I \mid\,<\,\infty.
\end{eqnarray}
Boundary conditions on the surface of the critical cylinder $x\,=\, 
x_a$ can be written as follows:  
\begin{equation}
T_{\mu}^{\nu}\,=\,{A}\,=\,{A}^{\prime}\,=\,0, \quad e^{\gamma} 
\,=\,1, \quad e^{\beta}\,=\,e^{-\mid C_2 \mid x}\,>\,0.
\end{equation}
The conditions (4.23) together with the relations $e^{2\gamma}\,=\,G\, 
{A}^2 \,+\,H,$ imply that $H=1.$  Therefore the metric (4.1) that
satisfies the regularity conditions reads:
\begin{equation}
ds^2 \,=\, (G {A}^2 +1) dt^2 - \frac{1}{(G {A}^2 +1)} \Bigl[
e^{2C_2x}\bigl\{dx^2 + d\phi^2\bigr\} + dz^2\Bigr].
\end{equation}
As in the previous case, we will study the system for different $P(I)$.

\noindent {\bf I.} Note that some class of regular solutions can be obtained
choosing $P(I)$ in the form
\begin{equation}
P(I) = P_0 (\lambda I - N)^s\,Q(\lambda I), 
\end{equation}
where  $Q(\lambda I)$ is some arbitrary, continuous, positive defined
function, having non-trivial value at the center; $\lambda$ is the 
coupling parameter; $N > 0$ is some dimensionless constant that is
equal to the value of $\lambda I$ at the center. The other constant
$P_0$ is defined from the condition $P = 1$ at spatial infinity
$\xi = 0$. For $R =$ const. one gets the most simple form of $P(I)$
that leads to regular soltions. As in the spherically-symmetric case, 
for the regular solutions  $\lambda \ge GN.$  

\noindent
{\bf a)} Choosing  $P(I)$ in the form
\begin{equation}
P(I) = P_0 (\lambda I - N)^2,
\end{equation}
we get
\begin{equation}
{A}(\xi) = \sqrt{\frac{N}{\lambda - GN}} \mbox{th}\,bx,
\end{equation}
where $b = \sqrt{C^2 N P_0 (\lambda - GN)}$, the integration
constant $x_1$ is taken to be trivial. The regularity condition implies
$b \ge 1.$ The metric function $e^{2\gamma}$, radial electric field and 
the total energy of the material field system can be written as 
\begin{equation}
e^{2\gamma} =  \frac{\lambda}{\lambda - GN}\Bigl[1 - \frac{GN}
{\lambda \mbox{ch}^2 bx}\Bigr],
\end{equation}
\begin{equation}
|{\bf E}| = |C| e^{\gamma -\beta} \sqrt{P(I)},
\end{equation}
\begin{eqnarray}
E_f = \frac{\lambda C}{2 G\sqrt{G}}\Bigl[\frac{\sigma}{\sqrt{1 - \sigma^2}}\,
\frac{\sqrt{1-\sigma^2}}{2}\mbox{ln}\,
\frac{1 +\sigma}{1 - \sigma} \Bigr],
\end{eqnarray}
where $\sigma^2 = GN/\lambda < 1.$ As one sees $|{\bf E}| \to 0$ as 
$x \to \pm \infty.$ The solution obtained satisfies all the regularity 
conditions and is a solitonian one. The density  of mass $(\rho_m)$ and 
the density of effective charge $(\rho_e)$ are
\begin{eqnarray} 
{\rho_m}|_{x \to -\infty} &\to& 
\,\left\{\begin{array}{cc} \mbox{const}& b =1; \\
                            0          & b >1; 
\end{array}\right. \nonumber\\ 
{\rho_m}|_{x \to +\infty} &\to& 0  \quad   b \ge 1; \nonumber\\
{\rho_e}|_{x \to -\infty} &\to& 
\,\left\{\begin{array}{cc} 2C^2\sqrt{G}(1-\sigma^2)/\pi\sigma & 
b =1; \\   0 & b>1; \end{array}\right. \nonumber\\
{\rho_e}|_{x \to +\infty} &\to& 0  \quad   b \ge 1. \nonumber
\end{eqnarray}
The total charge of the system is equal to zero.

\noindent
{\bf b)} Let us consider the case with $I_c = 0$, choosing
\begin{equation}
P(I) = \lambda I.
\end{equation}
In this case we get
\begin{equation}
{A}(\xi) = \frac{1}{\sqrt{G}\mbox{sh}\, (\sqrt{\lambda} C x)}.
\end{equation}
The metric function $e^{2\gamma}$ in this case reads 
\begin{equation}
e^{2\gamma} =  \mbox{cth}^2 \,(\sqrt{\lambda} C x),
\end{equation}
that gives
$$ e^{2\gamma}|{x \to \pm \infty} \to 1, \qquad 
e^{2\gamma}|{x \to \pm 0} \to \infty.$$
Inasmuch $e^{2\beta} = e^{-2\gamma + 2 C_2 x}$, $x = x_1 = -\infty$
corresponds to one of the axes of the field configurations. This axis is 
regular if $\sqrt{\lambda C} > 1$ and ${A} (x_1) = 0$ and 
$e^{2\gamma (x_1)} = 1.$ So for $e^{2\gamma}|{x \to \pm 0} \to \infty$,
one gets $e^{2\beta}|{x \to \pm 0} \to 0$, i.e. $x=x_2 =0$ corresponds to
the second, singular axis. In this case the solution obtained is defined
on $-\infty \le x \le 0.$ At $x \to + \infty$ 
$e^{2\beta}|{x \to + \infty 0} \to \infty$ and ${A}(x) \to 0$. It
means that $x = + \infty$ defines the spatial infinity. In this case
the solution is defined on $0 \le x \le \infty$ and possesses one singular
axis corresponding $x = 0$.

\noindent
{\bf II.} Let us now obtain the droplet-like configuration.
Choosing $P(I)$ in the form [see Figure 2]:
\begin{equation}
P(J)\,=\,J^{(1\,-\,2/\sigma)}\,\bigl[(1\,-\,J)^{1/\sigma}\,-\, 
J^{1/\sigma} \bigr]^2\,(1\,-\,J),
\end{equation}  
where $J\,=\, G\,I;\quad \sigma\,=\, 2n\,+\,1;\quad  n=1,2,3\cdots$  
one can find the expression for ${A}(x)$ which is similar to the one
in spherically-symmetrical case [see Figure 1]: 
\begin{equation} {A}(x)\,=\,(1/\sqrt{G})\,\bigl[1\,-\, 
\mbox{exp}\,\bigl(-\frac{2\,C\,\sqrt{G}}{\sigma}\,(x\,-\,x_0)\bigr) 
\bigr]^{\sigma/2}.
\end{equation}  
As one can readily see from (4.35), 
the conditions (4.22) and (4.23) are fulfilled if $\mid C_2 \mid 
\,\le\, C\,\sqrt{G}/\sigma.$  It is noteworthy that at $x\,\le\,x_0, \quad 
{A}(x)\,\equiv\,0$ and the space-time is flat, the gravitational 
field being absent \cite{ryb4}.  

\noindent
There is a principal difference between  solutions 
(3.37) and (4.35).  For the case of spherical symmetry the droplet-like 
solution can be transformed to the soliton-like one if  the boundary 
$\xi_0$ is removed by putting $\xi_0\,=\,0$ (as in this case $\mbox{exp} 
[\beta(\xi_0)]\,=\,1/\xi_0\,=\,\infty$).  On the contrary, for the case 
of cylindrical symmetry the removal of the boundary is equivalent to 
putting $x_0\,=\,-\infty,$ as in this case $\mbox{exp}[\beta(x_0)] \,=\, 
\mbox{exp}(-\mid C_2 \mid x_0)\,=\, \infty.$  Under this last choice the 
solution (4.35) takes constant value ${A}(x)\,=\,1/\sqrt{G}$ and  
the soliton structure disappears.  For the considered case, as  well as  
for that of spherical symmetry, the density of the field energy  is given 
by equation (3.23) and the  linear density of energy is similar to (3.24):
\begin{equation}
E_f\,=\,(C/4)\,\int\limits_{0}^{1/\sqrt{G}}\,d{A}\,e^{-3\gamma}\, 
\bigl[\sqrt{P}(1\,+\,e^{2\gamma})\,+\,4\,I\,(\sqrt{P})_I\,\bigr], 
\end{equation} 
Substituting $P(I)$ from (4.34) into 
(4.36), one can find that $E_f$ is finite and the total energy 
$E_f\,+\,E_g$ turns out to be zero.  

\noindent
Let us now define the effective charge density $\rho_e$ and total charge 
$Q$, corresponding to the unit length on z-axis. In generally from (2.4) 
one gets \cite{shik2} 
\begin{equation} j^\alpha\,=\,\frac{1}{4\pi} \, 
\bigl(\varphi_{, \beta}\,{\varphi}^{, \beta}\bigr)\, \Psi_I \, {
A}^\alpha, \end{equation} that for static radial electric field leads to 
\begin{equation}
j^0\,=\,\frac{C^2}{4\pi}e^{-2(\alpha+\gamma)}\, P_I\,{A}.
\end{equation}
Then for chronometric invariant electric charge density $\rho_e$ we have
\begin{equation}
\rho_e\,=\,\frac{j^0}{\sqrt{g^{00}}}\,=\,
\frac{C^2}{4\pi}e^{-(2\alpha+\gamma)}\, P_I\,{A}.
\end{equation}
The total charge is defined from the equality
\begin{equation}
Q\,=\,2\pi\int\limits_{x_a}^{x_\infty}\rho_e\sqrt{-^3 g}\,dx.
\end{equation}
Putting the corresponding quantities  into the foregoing equality after 
some simple calculations we obtain
\begin{equation}
Q\,=\,\frac{1}{2}\,e^{-2\gamma}\,{
A}^{\prime}\mid_{x_a}^{x_\infty}\,=\,0.\end{equation} 

\noindent
Now it is worthwhile to make again 
the comparison with the flat-space solutions of the equations (2.3) and 
(2.4), using  the interval:  
$$ ds^2\,=\, dt^2 \,-\,d\rho^2\,-\,\rho^2\, d\phi^2\,-\,dz^2.$$  
In this 
case the scalar field equation (2.3) admits the solution:
\begin{equation}
\varphi^{\prime}(\rho)\,=\,C\,P(I)/\rho, \quad P(I)\,=\,1/\Psi(I), \quad 
C\,=\,\mbox{const.}
\end{equation}
Inserting (4.42) into (2.4), one can 
find the electromagnetic field equation which admits the solution in 
quadratures: 
\begin{equation}
\int\limits_{}^{} \frac{d{
A}}{\sqrt{P(I)}}\,=\,\pm\,C\,\mbox{ln}\frac{\rho}{\rho_0},\quad
\rho_0\,=\,\mbox{const.} 
\end{equation}
Substituting $P(I)$
from (4.34) in (4.43), one gets  the solution of the droplet-like form:  
\begin{equation}
{
A}(\rho)\,=\,(1/\sqrt{\lambda})\,\bigl[1\,-\,\bigl(\frac{\rho}{\rho_0} 
\bigr)^{2\,C\,\sqrt{\lambda}/\sigma}\bigr]^{\sigma/2}.
\end{equation}
One concludes from (4.44) that ${A}(\rho\,\ge\,\rho_0)\,\equiv\,0.$ 
It means  that the electric charge of the system is zero . For 
the solution (4.34) the regularity conditions both on the axis $\rho\,=\, 
0$ and on the surface of the critical cylinder $\rho\,=\,\rho_0$ are 
fulfilled if $C\,\sqrt{\lambda}\,\ge\,\sigma.$  It  is noteworthy that 
in the case of cylindrical symmetry, both in the flat space-time and with 
account  of the  proper gravitational field, there do not exist any 
soliton-like solutions, as for the choice $\rho_0\,=\,\infty$ the 
solution (4.44) degenerates into the constant: 
${A}(\rho)\,=\,1/\sqrt{\lambda}.$  
The linear density of the field energy 
in flat space-time  can be found from the expression similar to (3.23), 
and as well as  in the case of spherical symmetry, it is equal to zero:
$$E_f\,=\,\frac{C}{2}\,\sqrt{P\,I}\mid_{{A}(\rho_0)}^{{A}(0)}
\,=\,0,$$ 
as was expected. 

\section{Discussion} \setcounter{equation}{0}

\noindent
Exact regular static spherically- and/or cylindrically-symmetrical 
particle-like solutions to the equations of scalar 
nonlinear electrodynamics in General Relativity have been obtained. 
As a particular case we found a class of regular solutions
with sharp boundary (droplet-like solutions or simply droplets). It is 
shown that outside the droplet gravitational and electromagnetic fields 
remain absent i.e. total energy and total charge of the configuration are 
zero. We underline once more the principal difference between the 
droplet-like solutions with spherical symmetry and those with cylindrical 
one. In the first  case there exists a possibility of continuous 
transformation of the droplet-like configuration into the solitonian one 
by transporting the sharp boundary to the infinity. As for the second 
case, there is no such a possibility, and the soliton-like configuration 
disappears when the boundary is smoothed tending to the infinity.  Further 
we intend to study the interaction processes of droplets with 
external electromagnetic and gravitational fields and also the scattering 
of photons and electrons on droplets.

\vskip 10mm

\noindent
{\bf Caption of figures}

\vskip 3mm
\noindent
Figure 1. Perspective view of droplet-like solution.  
The configurations are plotted for $\lambda = 1, \quad x_0 = 2$ and 
$\sigma$ takes the values $3,\ 5,\ 7,\ 9$. (Note that in the 
figures illustrated here $3,\ 5,\ 7,\ 9$ correspond to the value 
of $\sigma$ ).

\vskip 3mm
\noindent
Figure 2. Perspective view of the inverse function to 
the interaction one (i.e. $P(I)$) that provides us with the droplet-like 
configurations (Figure 1). As is seen from Figure 1, the stronger the 
interaction  the more localized the corresponding droplet-like configuration. 
\end{document}